\documentclass[aps,prd,superscriptaddress,eqsecnum,amsfonts,showpacs,epsf]{revtex4}
\usepackage{epsfig}

\newcommand{\be}{\begin{equation}}
\newcommand{\ee}{\end{equation}}
\newcommand{\bea}{\begin{eqnarray}}
\newcommand{\eea}{\end{eqnarray}}

\newcommand{\nn}{\nonumber}

\begin{document}


\title{Pions and excited scalars in Minkowski space DSBSE formalism}

\author{V. \v{S}auli}

\email{sauli@ujf.cas.cz}
\affiliation{Department of Theoretical Physics, Institute of Nuclear Physics Rez near Prague, CAS, Czech Republic  }

\begin{abstract}

We present the  solution of Schwinger-Dyson and  Bethe-Salpeter equation (BSE) for a light 
flawour non-singlet spinless meson in Minkowski space.
The equations are solved in momentum space without the use of the  auxiliary Euclidean space.
To exhibit that Minkowski space non perturbative calculations are actually possible is the main purpose of presented paper.
According to confinement, the quark propagator is regular in momentum space, which  allows us to look for  the BSE solution directly in the Minkowski momentum space. We find the Minkowski space solution for confining theories is not only numerically accessible, but also provides a  reasonable description of the  pseudoscalar meson system. However, in the case of scalar mesons, the model is less reliable and we get more extra light states  then observed experimentally.

\end{abstract}

\maketitle


%
\section{Introduction}

Understanding of hadrons represents difficult tasks within the use of QCD degrees of freedom -quark and gluon fields. In a more or less effective   approaches including e.g. QCD sum rules, effective chiral Lagrangians, effective Heavy quark methods, Hamiltonian light-cone approach, ADS/CFT correspondence etc. the problem is circumvented by an additional, not always obviously satisfied,  assumptions. On the other side,  Lattice QCD as well as functional Schwinger-Dyson equations (SDEs)  in Euclidean space approximation (EA) are a conventional tool based on non-perturbative utilizing of quark and gluon degrees of freedom.   

The last mentioned approaches should  be equivalent when solved exactly and when vanishing lattice spacing is achieved, at least when the both are defined in the auxiliary Euclidean space. For decades, it is largely  believed that the hadron spectra and related form factors for the timelike momenta should be feasible by an analytical continuation of the real data originally calculated and collected in the Euclidean space. Recall, the region of timelike momenta is just  where the resonances $\rho(n),\psi(n), Y(n)...$  exhibit variety of the peaks in a various form factors.    
In practice, theorists have longstanding tremendous obstacles with an analytical continuation of Euclidean results to the  Minkowski space.
Until now,  neither the lattice nor the SDEs provide a form factor or cross section for the continuous timelike arguments. 
The lattice has its on problems with the inclusion of the light quarks, while the EA  Dyson-Schwinger-Bethe-Salpeter equations (DSBSEs)  calculations show an incredible troubles due to the appearance of (complex conjugated) singularities when  QCD Green functions are evaluated at complex argument. For instance, in  recent DSBSEs treatment, this lead to the overlap of error bars for a different radial excitations in the light sector \cite{FKWa2014} as well as in the sector of mesons made up from heavy quarks \cite{HPGK2014,FKWb2014}. A missing knowledge of propagators at complex momenta makes a precise identification of energy levels impossible. Contrary,  (not only) recent experiments do the best in the timelike region, the electroproduction of hadrons represent one of the most precisely measured QCD process: the pion form factor shows up unambiguous presence of rho like resonances, also five or six vector meson resonances are known for charmonium and bottomonium  with increasing energy of $e^+e^-$ pair.

In presented paper, the author ignores the conventional wisdom and does not follow  historically prior suggestions and do not solve the DSBSEs in the Euclidean space, but instead of, the calculations are performed directly in Minkowski space.  
This somehow numerically  inconvenient way, however leads to the results, which obviously are not  an analytical continuation of the Euclidean theory.
The Greens functions become complex valued at the regime where the Euclidean counter-partners must be real from the definition. 
Actually,  we argue that this is phenomena of  confinement, which is responsible for such behavior and which also makes DSBSE system soluble  directly in the Minkowski momentum space. This simple fact has been overlooked or perhaps ignored by a community and to the  author knowledge, the first Minkowski space BSE solution has been  shown  for charmonium system \cite{SAUetac} only very recently. At this place one should also mention other Minkowski space BSEs solutions \cite{KUSAWI1997,SAUADA2003,SAU2008,CARKAR2010,FRSAVI2012,CARKAR2014}. These methods  are based on Perturbation Theory Integral Representation (assuming validity of usual Wick rotation and positivity),  which  perfectly fit for a weakly bounded system of particles. These methods do not contradict with EA results,  however they are not suited for confining theory and they would require some nontrivial extension in order to provide a correct description of QCD hadrons.

The quark and the gluon propagator and their vertices are building blocks of equations for hadrons and their knowledge is thus necessary 
for the solution of DSBSEs. Usually one first solve gap equation for quarks (and gluons if possible) and then one look for the solution of Bethe-Salpeter
equation for meson or for the relativistic Fadeev equations for baryons. A small feed back of bound states is expected and such modification of GFS has been studied in EA DSBSEs  \cite{FIWI2008}. In Minkowski space, one has to overcome the problem with not well behaving numerics. In many  respects  incomplete, however exciting  minirewiev of Minkowski space solutions for a quark propagator has been presented in the author's paper \cite{SAULIlat}. Very generally, the solutions are complex everywhere and if the effective coupling is strong enough the propagators do not show the real poles -the quarks and gluons do not  move as a free single particle with definite mass. Furthermore, depending on the model details  one can observe two  type of the solutions:   oscillating  and the one which is more smooth (noting , there is no sharp borderline in between). 
From solely known solution of gap equation, it is quite impossible to decide, which scenario is more realistic, the calculation of meson spectra can give the first hints. In the presented paper  we restrict ourself to the second case  here, leaving the fluctuating solutions for future comparative study.
The  DSBSEs  model for pions and isovector scalars are considered here.

\section{Rainbow-Ladder SDE-BSE model}

The pion is well established pseudo-Goldstone boson of broken chiral symmetry in QCD.
As a consequence of a small current quark mass,  it is not completely massless, but very light. 
The first pionic excitation is as heavy as the other excited mesons, providing thus the ratio of pionic intercept is relatively largest  among the all mesons. Further property of pion is connected with the structure of pseudoscalar $\Gamma_5$ and pseudovector $\Gamma_5^{\mu}$ vertices, which
are related related through the axial-vector Ward-Takahashi identity

\be \label{AWT}
P_{\mu}\Gamma_5^{\mu}(k,P)=S^{-1}(k_+)\gamma_5\frac{\tau^{j}}{2}+\gamma_5\frac{\tau^{j}}{2}S^{-1}(k_-)
-M(\mu)\Gamma^j_5(k,P)-\Gamma^j_5(k,P) M(\mu)
\ee
where ${\cal S}=diag[S_u,S_d]$ , ${\cal M}(\mu)=diag[M_u(\mu),M_d(\mu)]$ are dressed quark propagator and current quark mass matrices and the arguments satisfy $k_+ +k_-=P$  $k_+ -k_-=k$.
The propagator satisfies the SDE (the gap equation), which reads
\bea  \label{quarkgap}
S^{(-1)}&=&\not p -m_o-\Sigma(p)
\nn \\
\Sigma(p)&=&i\int\frac{d^4k}{(2\pi)^4} G_{\mu\nu}(p-k)\frac{\lambda^a}{2}\gamma_{\mu}S(k)\Gamma^a_{\nu}(p,q) \, ,
\eea
where $M$ in $(\ref{AWT})$ is the running quark mass at the scale $\mu$, $\Gamma^a_{\nu}(p,q)$ is the quark-gluon vertex satisfying its own SDE and 
$G_{\mu\nu}$ is gluon propagator.

In narrow mass approximation the axial-vector and pseudoscalar vertices exhibit poles for $P^2=m^2_{\pi_n}$, where $m_{\pi_n}$ is the mass of the pion 
(for $n=1$) or of the arbitrary excitation $\pi(1300)$, $\pi(1800)$,... .

Projecting out the bound states from inhomogeneous equation for axial vertex one can get
the bound states BSE for pion:
\be \label{mesonbse}
\Gamma^j_{\pi_n}(p,P)=i\int\frac{d^4k}{(2\pi)^4}\chi(k,P)^j_{\pi_n}K(k,p,P)
\ee
where $K(k,p,P)$ is quark-antiquark interaction kernel and 
where the total momentum satisfies $P^2=m^2_{\pi_n}$.The bound state pseudoscalar BSE vertex function $\Gamma^j_{\pi_n}(p,P)$ as well as the BSE  wave function $\chi(k,P)=S(k_+)\Gamma(k,P) S(k_-)$ are fully determined 
by the four scalars, the decomposition reads
\be
\chi(k,P)^j_{\pi_n}=\tau^j\gamma_5\left[E_{\pi_n}(k,P)+\not P F_{\pi_n}(k,P)+
 \not k G_{\pi_n}(k,P)+[\not k,\not P] H_{\pi_n}(k,P)\right].
\ee
The residue of the axial vertex function at $P^2=m^2_{\pi_n}$ are determined by pseudoscalar meson leptonic decay constant.  

The homogeneous BSE  is valid for any spin meson (\ref{mesonbse}), the decomposition for a parity plus scalars reads
\be
\chi^j_{a_n}(k,P)=\tau^j\left[E_{a_n}(k,P)+\not P F_{a_n}(k,P)+
 \not k G_{a_n}(k,P)+[\not k,\not P] H_{a_n}(k,P)\right].
\ee

In order to get a correct  electromagnetic form factor and pion charge radius   a symmetry preserving truncation of infinite DSBSEs system is required. The simplest electromagnetic Ward-identity preserving truncation is the  so called rainbow-ladder (RL) approximation, which we are going  
to use in our Minkowski space model as well. 
From this it follows, that the  kernel used in the quark SDE must enter the BSE  for mesons as well. For the status of solution of DSBSEs system in EA see for instance \cite{FKWa2014,HPGK2014,FKWb2014} and references therein.

\section{State of art of Minkowski space solution}

To complete the model we use the  kernel already introduced  in the paper    \cite{SAULIlat}
solely for purpose of study of quark propagator behavior.

The rainbow-ladder approximation enables us to write down the effective charge $\alpha$:
\be \label{gabii}
\Gamma^{\mu}(k,p)G_{\mu\nu}(k-p)\rightarrow \gamma^{\mu}  G_{\mu\nu}^{[free]}(k-p) \alpha(k-p)
\ee
defined as a simplified product of the quark-gluon vertex and the gluon propagator. The  BSE and DSE kernel
introduced above can be related with the charge $\alpha$ in the following way:
\be \label{KS}
K(x)=\frac{\alpha(x)}{\pi x}=(8\pi^3)C \lambda^2 \frac{e^{-\frac{\sqrt{-x}}{\lambda}}\Theta(-x)+e^{i\frac{\sqrt{x}}{\lambda}}\Theta(x)}{(x-\lambda^2)^2+\lambda^4} \, ,
\ee
 where $k,p$ are momenta of incoming and outgoing quark respectively. 
The effective charge depends on a single variable: gluon four-momentum $x=q^2$, which is a major simplification 
when comparing to the full momentum dependence of l.h.s. of Eq. (\ref{gabii}).
In the expression (\ref{KS}) a letter $C$ stands for an effective coupling strength and $\lambda\simeq \Lambda_{QCD}$ is a setting scale.

In QCD the quark propagator has the general form 
\begin{figure}[t]
\begin{center}
\centerline{  \mbox{\psfig{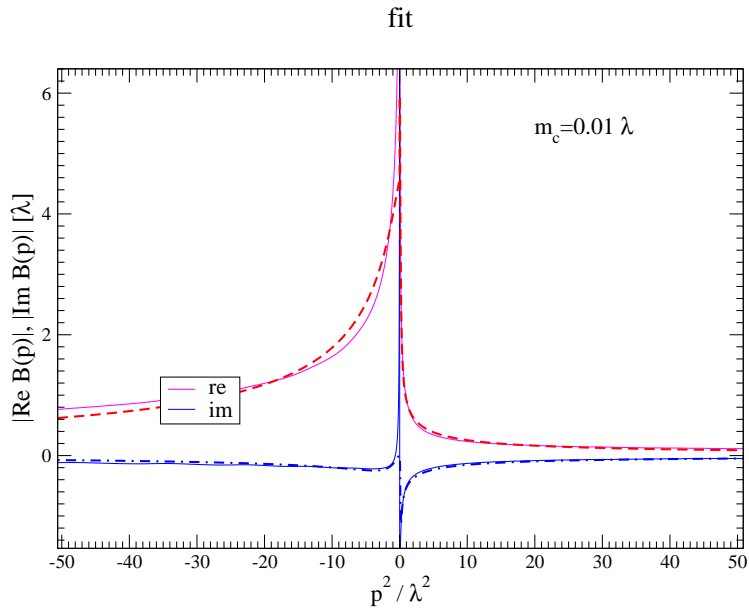}} }
\caption{Solution of QCD gap equation: 
Minkowski space complex selfenergy -the quark dynamical mass function- and  the analytical fit (\ref{fit}) are displayed.
\label{firstpic}}
\end{center}
\end{figure}

\be \label{SDE}
S^{-1}(p)=\not p A(p^2)-B(p^2)=1/Z(p^2)\left[\not p-M(p^2)\right] \, ,
\ee
where two scalars $A,B$ are determined by the SDE (\ref{quarkgap}). It is well known from EA, that the mass function $M$  bellow $\Lambda_{QCD}$ scale
is blowing rapidly up (at least) to the constituent quark mass value $M\simeq \Lambda_{QCD}$ and even infrared divergence \cite{AGNEPA2005,NEPA2006} in the origin is not excluded in general.

Explicitly we get for the function $B$   
\bea \label{defker}
B(p)&=&\frac{Tr}{4}ig\int\frac{d^4k}{(2\pi)^4} \gamma^{\mu} t^a S(k)\gamma^{\nu} t^a G_{\mu\nu}^{[free]}(q)\alpha(q)
\nn \\
&=& i\int \frac{d^4k}{(2\pi)^4} \frac{B(k)}{k^2-B(k^2)} K(q^2)
\eea
in RL approximation. Albeit not guaranteed in Minkowski space, we  neglect  momentum dependence of the function $A$ by simply taking $A=1$, which is known to be reasonable choice in Landau gauge in EA. Anticipate already in this paper, that the inclusion of a correct UV behavior in the kernel $K$ (i.e. free, log modified one gluon exchange) is numerically accessible  as well, however here  we prefer to leave related solution and  discussion for an incoming study \cite{SAULI-II}. 

Recall , that in order to get nontrivial solution one typically needs $(8\pi^3)C \simeq 1$, 
irrespective of presence or absence of perturbative  ultraviolet gluon.
We work in Landau gauge for convenience, noting that the factor $3/4$ arising after the projection is exactly absorbed
by  Casimir of $SU(3)$ color group.   Let us already now anticipate that the  coupling $C=1/90$, which corresponds 
with the one of the  ladder models considered already in the paper \cite{SAULIlat} implies a relatively large scale $\lambda=0.985 GeV$.

In order to  get the solution, one first solves the gap equation (\ref{defker}), whose solution is required to complete BSE kernel, and then solve the BSE for pion and all its radial excitation. In order to get the first hint for the pion spectrum, we consider only the first component of the BSE vertex function, i.e.  
\be \label{approx}
\Gamma(k,P)_{\pi_n}=\gamma_5 \Gamma_{E_{\pi_n}}(k,P)
\ee
for the pseudoscalar and for the scalars we retain only the term $\Gamma_{E_{a}}(k,P)$, recall they were enough to get ground state in almost any known EA.  

A single real pole related with  on-shell  mass, an induced branch point singularities associated with various production thresholds, they are attributed to physical free particle motion and they  are not featured in the quark and the gluon propagators in  Minkowski space.  
In accordance to confinement, they are absent due to the strong coupling. In other words,   all particle-like  singularities  are moved from real axis somewhere into the complex plane of square of momenta. It does not mean that the real axis singularity is excluded at all, however  its character is not associated with a particles. In fact,  an early  hints were already coming  with  Euclidean studies, and as it was repeatedly found,  an analytically continued  quark and gluon
 propagators had a multiple complex conjugated branch points \cite{STACAH1990,STINGL1,MAR1995,STINGL2} there. More generally, the lost of positivity, which is equivalent to the absence of  Khallen-Lehmann representation, is generally  expected to be in accordance with confinement. 
Important for us is the technical point, since all these things turn the Minkowski space quark and  gluon propagator more smooth   for a real valued momenta. This is the main reason which  makes the solution of DSBSEs numerically feasible in Minkowski space.

Numerical integrations 'over the hyperbolas'
\be
k^2=g^{\mu\nu}k_{\mu}k_{\nu}=k_0^2-{\mbox \bf k}^2 \, .
\ee
dictated by Minkowski metric, turns a numerical search to longstanding  game. We did not even attempt to solve the  BSE by  matrix inversion method, which is common procedure in the Euclidean space otherwise. Instead of this, we use the method of iterations, which has been found successfully operating  for BSE in EA already, as well as for the charmonium BSE in Minkowski space \cite{SAUetac}. The BSE is solved in the rest-frame of the meson $P=(M,{\mbox \bf 0})$, thus the Lorentz invariant arguments of BSE $p.P=p_0 M$ and $p^2$  are related in an easy way with the pair $p_o, p^2$ which was chosen numerically for the integrations.

The BSE can be immediately written into the form

\be
\Gamma_i(p_0,p^2;P)=i\Sigma_j\int_{-\infty}^{\infty} dk_o \int_{-k^2}^{\infty} dk^2 \sqrt{k^2-k_o^2}\int_{-1}^{1} dz f_j(k,^2,p^2,k.p,k.P) \, \, ;
\ee
where $z$ is cosine of the angle between internal and external spacelike momenta and 
where $f$ represents the product of scalar functions, which arises after a various  pojections of the matrix $S{(k_+)}\Gamma(k_o,k^2;P)S(k_-)K(k,q,P)$.
Due to the Poincare invarinace one can simply take $p=(p_0,0)$ for the timelike and $p=(0,p_z)$ for the spacelike relative momenta, which allows further reduction of  the BSE to two-dimensional integral equation at least at the timelike region. In isospin limit (equivalent to equal mass case for particle constiunet case) we further reduce number of points for $q_0$ assuming $\Gamma$ is an even function with repect to $k.P$, and we take
\be
 \int_{-\infty}^{\infty} dk_o  f(k_o.p_o,k_o,...)= \int_{0}^{\infty} dk_o \left[ f(k.p=k_o.p_o,k.P=k_oM,...)+f(k.p=-k_o.p_o,k.P=-k_oM,...)\right]
\ee
and $p^2=p_0^2$.

For a spacelike argument we integrate over spacelike cosine $z$ numerically, finding that relatively large number of integration points is required to get rid of various fake solutions. The gluon momentum $q^2$ in ladder kernel is then  given as following  
\be
q^2=(k-p)^2=k^2+p^2+2\, \sqrt{k_o^2-k^2}\, p\, z
\ee
for $p=\sqrt{-p^2}$, while for the timelike external argument it has been  choosen  as
\be
q^2=(k-p)^2=k^2+p^2-2 k_o\, p
\ee
for $p=\sqrt{p_o^2}$. For the numerical integration, it is actually useful to take the grid of variable $k_0$ as square root of the positive  $k^2$.

 Gaussian integrator is used providing  several hundred  milion  cells of discretized Minkowski 3+1 dimensional momentum  volume $(N_{DSE/BSE}\simeq few 10^3/10^2$ for $k^2$ and  $k_0$, and $N_z=80-160$ was the number of  point for spacelike  integration cosine.  In the case of BSE, for  each meson mass $P^2$, a  several iterations was performed then $P^2$ was gradually changed by a numericaly impercetimble small step in order to prevent a lost of convergence. I use auxiliar "eigenvalue" function $\tilde{\lambda}$ exactly in a way it is described in the previous two studies \cite{SAUetac}. It has been introduced in BSE in the following manner: 
\be
\tilde{\lambda}\Gamma=i\int \chi K 
\ee
Then the solution is identified whenever $\tilde{\lambda}=1$ and the error $\sigma$ -a weighted difference between two consequtive iterations-   is minimal and vanishing. To avoid a fake solutions one should requirea a trajectory $\tilde{\lambda}(P)$ is smooth enough, which is nontrivial condition to be satisfied numerically (there are no obvious rules to achive a stability and one has to find  a stable numerics by a random testing. Most of  numericaly succesfull codes are published at the author web pages).   
  
The quark propagator needs to be evaluated for various arguments, say at $p_{\pm}$ as appears in BSE kernel. The region of the integration variables is thus restricted to a real domain. One does not need to continue to  complex arguments  at all, which is a clear advantage when compared to EA,
 where an obstacles arise especially for highly excited states. On the other hand one can observe increase of numerical noise for quark propagator at high $k^2$. I order to  eliminate one of the sources of potential numerical instability we avoid the  inter/extrapolations and the following fit is made for the quark propagator:
\bea \label{fit}
{\mbox Re} \frac{M(p)}{\lambda}&=& \frac{8a}{(a-b)^2+b^2}\log{(1-a)}+m_0
\nn \\
{\mbox Im} \frac{M(p)}{\lambda}&=&\frac{q^2}{(1-x)^2+1}\log{(1-x)} \, ,
\eea
 where  $x=q^2/\lambda^2$,  $a=x-1$, $b=0.1$, for $q^2<0$ . While for the  timelike argument $q^2>0$  the fit, which has been actually used reads
\bea
{\mbox Re} \frac{M(p)}{\lambda}&=&\frac{u(1+u)\log{(1+u)}}{x^2+v^2}+m_b
\nn \\
&+&\frac{\log{(3/2+u)}}{10 (x^2+v^2)}
\nn \\
{\mbox Im} \frac{M(p)}{\lambda}&=&-\frac{5 x \log{(1+x)}}{8 \left[(x-v)^2+v^2\right]}
\eea
,where for a now $v=0.1$ $u=v+x$.

For an achieved accuarcy of the selfenergy fit, see fig. \ref{firstpic}. The complex mass function has  vanishing imaginary part, however is complex even for the real argument everywhere, while the real part has its perturbative limit given by UV asymptotic  $m_b^{u,d}\rightarrow 0.01 \lambda$.
As a consquence the quark propagator functions
\be
\frac{\not p}{p^2-M^2(p)} \, \,; \, \, \frac{M(p)}{p^2-M^2(p)}
\ee
are definitely regular at the integration domain.

\begin{figure}[t]
\begin{center}
\centerline{  \mbox{\psfig{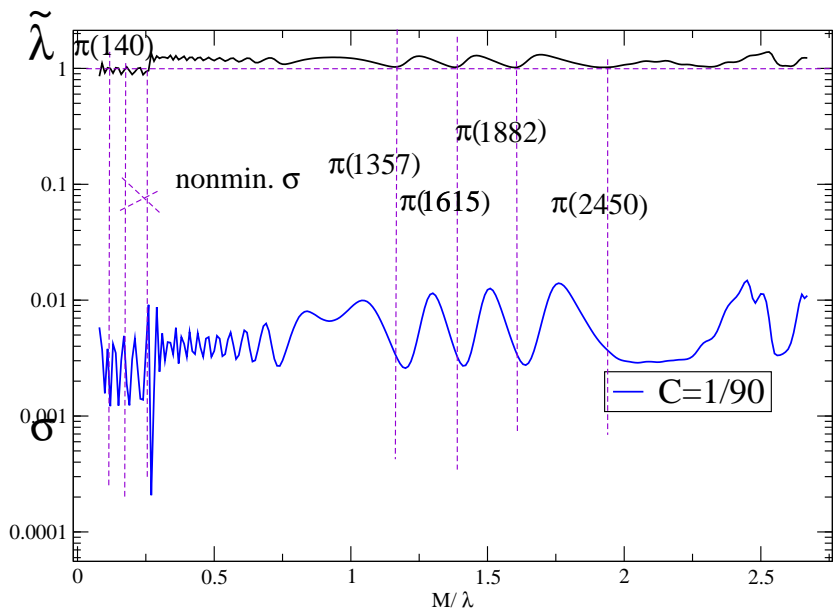}} }
\caption{Example of numerical search for pseudoscalar, the eigenvalue $\tilde{\lambda}$ and the weighted  error $\sigma$ are shown. 
Vertical lines label where the solutions are, while  crossed vertical lines represent quasisolution fakes. 
\label{pion22}}
\end{center}
\,

\mbox{  }

\,
\end{figure}

\section{Results and discussion}

Up to one extraordinary state,  for the value $C\simeq 1/90$ of the effective coupling, we  got a reasonable  agreement of obtained pionic states with an experiments. one can find, that the position of lowest lying state -the pion- is not  too much sensitive  with respect to a changes in  $C$, however this is the intercept which has to be tuned by the effective coupling. The evolution of $\tilde{\lambda}$ with $P^2$ is collected and shown in Fig.
\ref{pion22} for $C=1/90$. The all bound states can be identifien im, noting the data presented would require eight weeks run on a single (3.5GHZ) processor machine.   
\begin{figure}[t]
\begin{center}
\centerline{  \mbox{\psfig{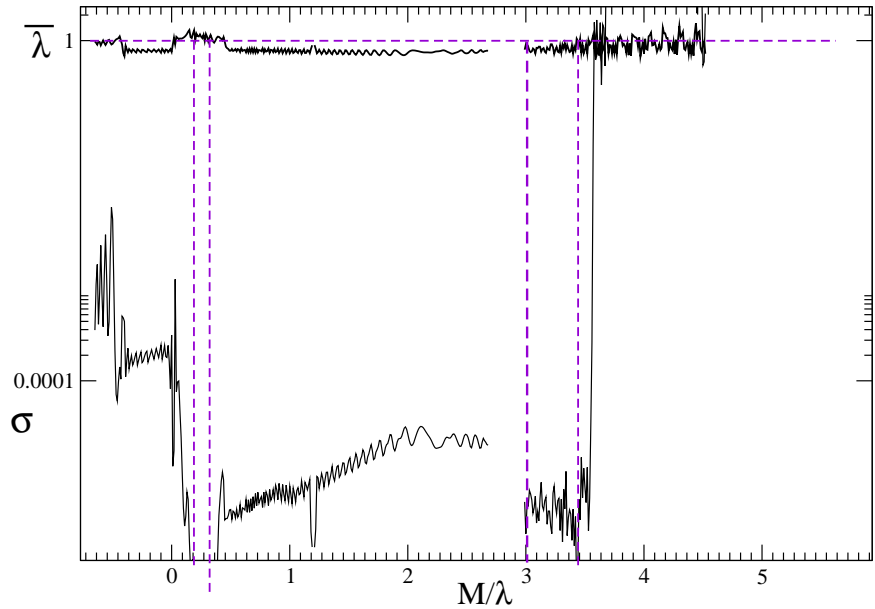}} }
\caption{Example of numerical search for scalars for $C=1/90$, the eigenvalue $\tilde{\lambda}$ and the weighted  error $\sigma$ are shown. 
Vertical lines label where the physical solutions are.\label{pokus2}} 
\end{center}
\end{figure}

It is matter of the fact that one get two,  instead of one, very light states for the coupling strength $C=1/90$, say at the value $0.115\lambda$ and $0.145\lambda$. The eigenvalues and errors  $\tilde{\lambda}, \sigma$   exclude that  one of them is a numerical quasisolution or fake. 
To get rid of the extraodinary state one has to note that a deccrease of the effective coupling leads to the growth of number of infrared bound states and vice versa. Thus for instance, for the value $C\simeq 1/50$ we get just one light bound state, however,  the same coupling  underestimates the masses of excited states as it would lead to completely wrong intercept. Going in opposite direction, i.e. decreasing slightly the coupling a new extrastates arise. The only correct conclusion is that LRA is inafficient for description of the pion. At least one has to go beyond 
 running coupling approximation as actualy  simple introduction of $P^2$ dependence of  the kernel  solve the problem. Here, in our case 
a simple theta function, which separate between ground state and excited states, is enough to achive an agreement with the experimental. It leave us with lighter state -the true pion-,  whilest a convergence is not achieved for the heavier solution and it vanishes for larger $C$.
We therefore argue that correct description of all mesonic states lies beyond  LR  truncation of DSEBSEs system, however a required changes in the kernel acould be only marginal  (and let me note, that taking into account one loop skeletons, which could lead to such desired corrections e.g. crossed box diagram in Minkowski space, is far beyond today computational ability).

\begin{figure}[t]
\begin{center}
\centerline{  \mbox{\psfig{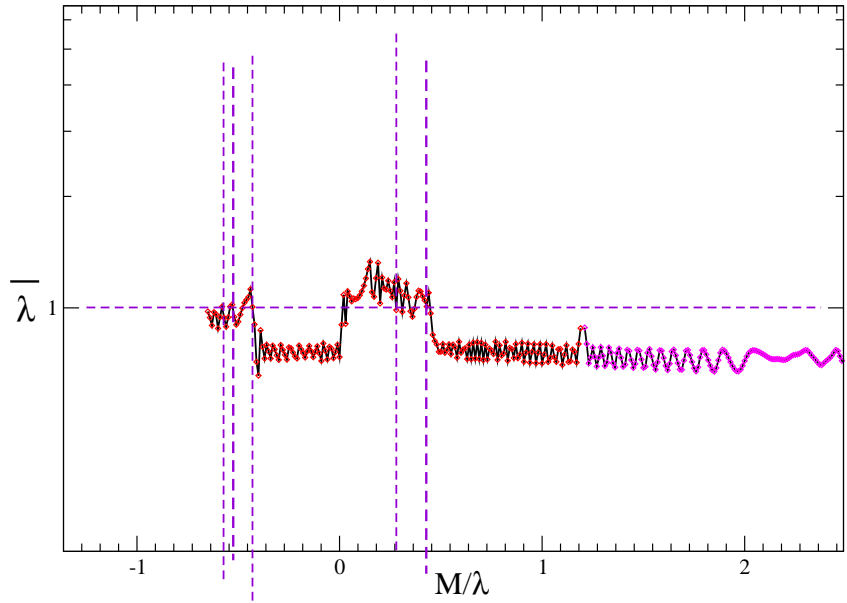}} }
\caption{Example of numerical search for scalarsfor $C=1/90$, the eigenvalue $\tilde{\lambda}$ and is shown in the infrared. 
Vertical lines label where the solutions are. Tachyonic solutions are displayed  as a curosity. \label{pokus}}
\end{center}
\end{figure}
There is an even worse situation in the case of  scalars, where for instance for $C=1/80$ we observe two or three states in the interval $300- 400 MeV$ (assuming $m_{pi}/\lambda$ =0.115 is the correct scaling) and four or five states for $C=1/90$ in the interval $400- 600 MeV$, recall a high accuracy achieved since $\sigma\simeq 10^{-8}$ in the later case. For the whole results  see fig. \ref{pokus2} , while the details of eigenvalue evolution are visible in fig.\ref{pokus}.  Such pattern is not observed in nature, and the first experimentally observed  scalar candidate is as heavy as $1GeV$ with radial excitations  basically  unknown. Recall also that neither of calculated states  can be directly related with the well known $\sigma$ resonance, which is very likely isoscalar and hence the interaction is  different when compared to the pionic case.  The isoscalar kernel includes two gluons aninhilation and unavoidably the admixture with  gluonic content is allowed. However for scalars, the  extragluonic contribution is not anomalously enhanced as in the case of pseudoscalar $\eta$ and one can expect large degeneracy between flawour singlet and nonsinglet mesons.
  Recall also here, that the observed broad resonance $\sigma(500)$  is tentatively explained as a isoscalar tetraquark excitation or tightly bounded pionic molecule \cite{CACOLE2006}.  It further exhibits itself in $J\Psi$ decay as meassured at BES \cite{BES} and three pions decay of $D^+$ as well meassured in E791 \cite{E791}. Regarding isoscalar $sigma$, it is believed that mixing with $qqqq$ slightly lower its mass as has been predicted from Coulomb gauged effective Hamiltonian \cite{CGWS2008}. In any case, the most accepted interpretation of sigma as a tetraquark resonance does not exclude that $qq$ state  survives as a mixing component with approximately similar mass and broad width. We speculate here, that numerical observation of several states made in our paper, can be a consequence of a bad narrow approximation atributed to the use of homogeneous BSE. 
 
 At this place, it is worthwile meaning, that the models based on EA of BSE \cite{FKWa2014,FKWb2014}  typically predicts one state at vicinity of $\sigma$, arguimg in adition that the lightness of the state is atributed to LR approximation, which is inadeuqate in the scalar channel.  
Our model predicts no heavy scalar states until $3 GeV$, where  just the one heavy scalar excitation is seen from our BSE solution. This is hardly experimentaly observable and we do not infer any strong conlusion from this solution.   
For completeness recall here, that for scalar quark-antiquark reonance, there is a null theorem \cite{scadron1, scadron2} providing why $qq$ component of $\sigma$  can be invisible in various process. These are the cancelation between nonresonant closed quark boxes and resonant graphs which includes quark trinagles with the $\gamma_5$ (two pions), which gives zero at $p\pi_-\rightarrow n\pi^+\pi^-$ amplitude measured at Crystal Ball collaboration \cite{CRYSTAL}.

\section{Summary}

The excited states of the scalar and the pseudoscalar mesons have been calculated   within DSBSEs formalism. In addition to almost massless pseudoGoldsotone 140 MeV pion, we have found its four excited states as well, giving 2450 MeV for the heaviest one.  
We get the solution for scalars as well, obtaining several and  rather close solutions at the range of $2m_{pi}$ or  $\sigma$ mass. 
Obviously, the light isovector Lorentz scalar meson is not observed yet in the experiments and in fact even the heavist candidate would be as twice lighter as $a(980)$. The scalar sector is not easily explained by QCD degrees of freedom and also presented Minkowski space BSE model fails by predicting the states which are not observed. 

 The  main purpose of presented paper was to present the first  Minkowski space nonperturbative calculation of bound state equations for a light mesons. This was possible due to to the special property of Green's functions describing confined quarks and gluons. Although the approximations of DSBSes in Minkowski space are limited by numerical convergence the presented model is an example, where working directly in Minkowski momentum space gives accurate and reliable reults. It does not require (a numerically sometimes imppossible) analytical continutation of the data coming from an auxilliary Euclidean space. 

If one is willing to accept the scenario proposed an unusual conclusion is evident: the Green's functions are complex objects in Minkowski space, but not only above certain timelike scale but everywhere. This is in agreement with absence of quark pole masses, quark production threshold and Lehman representation for the quark propagator.  It is plausible, that chiral symmetry breaking and confinement is intimately related with non-particle like singularities and hence with unusual complexity of Greens functions.
The  Green's function are complex in the timelike domain as well as in the spacelike, the later property explicitly exhibits intrinsical inequivalence of quantum field theory with Euclidean and Minkowski metric as definite ones. While the author generaly believe that EA must be a good one, the difference from  ussage of diferent metrics (and the similar approximation in DSE system) in different spaces can be expected, with nontrivial conquences, not only in an unubservable confined sector but for observable one. The comparison of calculated spectra and form factor will remain the only way to discriminate, what will be better in given cimcurstances. While I expect a  numerical obstacles will allways remain in Minkowski space, and  many approximations already known and studied in EA will be unavailable, however the ones giving fruits, should be extremelly usefull  for calculation of procesess with the timelike arguments. 

As an Euclidean Quantum field theory differ from the Minkowski one, it is not surprising that the analytical structure of both theories differ from each other. One should not expect that the analytical continuation, although it is meaningfull, well defined and feasible procedure is some cases, could necessarily provide the experimental data collected in Minkowski space.        

We have not yet calculated $f_{pi}$ from its definition since a quantitative study would require a self-consistent solution  for the quark propagator and BSE employing fewer approximations, e.g. the neglected components of BSE vertex should be considered as well. However, just for a ground state pion, one need not solve the complete pseudoscalar BSE, since the leading term $\gamma_5 B(p^2)/f_{\pi}$ in the limit $m,P\rightarrow 0$ is completely given by solution of corresponding DSE. Comparing normalized BSE and DSE solutions  gives us  $|f_{\pi}|\simeq (90 MeV )^2$.


%
\end{document}